\documentclass[11pt]{article} 

\usepackage[utf8]{inputenc} 

\usepackage[margin=1in]{geometry}
\usepackage{graphicx} 
\graphicspath{ {./images/} }

\usepackage{booktabs} 
\usepackage{array} 
\usepackage{paralist} 
\usepackage{verbatim} 
\usepackage{subfig} 
\usepackage{xcolor} 
\usepackage{mathrsfs} %

\usepackage{fancyhdr} 
\usepackage{sectsty}
\allsectionsfont{\sffamily\mdseries\upshape} 

\usepackage[nottoc,notlof,notlot]{tocbibind} 
\usepackage[titles,subfigure]{tocloft} 

\usepackage[hidelinks]{hyperref}
\usepackage{amsmath,amsthm}
\usepackage{amssymb}

\newtheorem{proposition}{Proposition}

\title{Generic Naked Singularities in Vaidya Spacetimes}

\begin{document}

\author{}
\date{James Wheeler}
\maketitle

\begin{abstract}
We investigate the occurrence of naked singularities, local and global, in the incoming Vaidya spacetimes with zero initial mass. While it is well-known that these spacetimes admit locally and globally naked singularities, we demonstrate that globally naked singularities are significantly more common than is stressed in the literature, being generic in a natural topology on this collection of spacetimes. A heuristic consequence of the results is that the slow accumulation of mass is both necessary and sufficient for both types of naked singularity in these spacetimes. We demonstrate that the naked singularity, as long as it exists, always has divergent curvature (Kretschmann scalar) associated to it along emerging null curves, regardless of the form of the mass function. In particular, a consequence is that the curvature strength of the singularity cannot be smoothed away.
\end{abstract}

\section{Introduction}

Among the largest outstanding problems in theoretical general relativity are the Cosmic Censorship Conjectures, which essentially posit that the theory is well-posed despite the assured phenomenon of singularities (\cite{hawking1965occurrence, oppenheimer1939continued, penrosesingularity, schoen1983existence}). These are a pair of conjectures, termed {\it weak} and {\it strong} (somewhat unfortunately, as technical formulations are independent of each other-- see, for example, Chapter 12 of Wald \cite{waldgr}), both originally due to Penrose (\cite{penrose1969gravitational,penrose1979singularities}). The weak version's heuristic content is that {\it dynamical singularities in general relativity are generically not visible to observers at infinity}, while the strong version's heuristic content is that {\it dynamical singularities in general relativity are generically not visible to any observer}. Singularities in violation of the weak version are dubbed {\it globally naked}, while those in violation of the strong version are dubbed {\it locally naked}.

While challenges in arriving at general proofs of comprehensive and compelling formulations of these conjectures abound, among the largest is nailing down precisely what {\it generic} is to mean in this context. Some such caveat is certainly required, as several examples of spacetimes containing naked singularities have been constructed (\cite{christodoulou1984violation, christodoulou1994examples, joshi1992strong, kuroda1984naked, rodnianski2019naked}). Arriving at the appropriate notion of genericness will doubtless require a thorough understanding of the extant examples. To aid this effort, we demonstrate here that one class of examples, the incoming Vaidya spacetimes, feature globally naked singularities which are much more prolific than is currently stressed in the literature.

It is well known that the incoming Vaidya spacetimes, perhaps the simplest possible models for the dynamical formation of a black hole, can also exhibit the dynamical formation of naked singularities (\cite{bengtsson2012spherical, ghosh2001naked, joshi1992strong, kuroda1984naked, lake1991naked}), i.e.\@ naked singularities to the future of a complete spacelike hypersurface (a nonsingular ``instant" in time). While the occurrence of the locally naked case has been sufficiently characterized in these sources, it seems the globally naked case has not been thoroughly explored outside of the highly restricted self-similar subclass and a few related examples. We treat both cases, as well as the singularity's curvature strength, in a uniform manner with simple ODE comparison techniques (distinguished from the ansatz approach of, say, Kuroda \cite{kuroda1984naked}, and the node analysis of Joshi and Dwivedi \cite{joshi1992strong}).

\section{Results}

We consider the spacetime  $M= (\mathbb{R} \times \mathbb{R}^3) \backslash \{(v,0,0,0) \, | \, v \geq 0 \}$ with metric given by 
\begin{equation}\label{equation:metric}
g=-\left(1-\frac{2m(v)}{r} \right) dv^2 + dv \otimes dr + dr \otimes dv + r^2 d\sigma^2,
\end{equation}
where $d\sigma^2$ is the metric on the unit sphere $S^2 \subset \mathbb{R}^3$, $r$ is the radial coordinate on $\mathbb{R}^3$, and $m: \mathbb{R} \to \mathbb{R}_{\geq 0}$ is nondecreasing, continuous, piecewise $C^1$, and satisfies $m(v) = 0$ for $v \leq 0$, $m(v) > 0$ for $v > 0$. That is, the mass parameter $m(v)$ (this is, e.g.\@, the Hawking mass of spheres of constant $r$ and $v$) increases continuously from $0$ beginning at $v=0$, with continuous derivative at all but finitely many points, at which left- and right-handed limits of $m'(v)$ exist and are equal to the left- and right-handed derivatives $m_\pm'(v)$. The Einstein tensor has a single nonzero component in these coordinates-- it can be computed to be
\begin{equation}\label{equation:curvature}
G = \frac{2m'(v)}{r^2} dv^2,
\end{equation}
from which we easily deduce that the dominant energy condition (DEC) is satisfied since $m'(v) \geq 0$. Strictly speaking, this observation is only entirely unambiguous when $m(v)$ is globally $C^2$, but the results which follow hold for the piecewise $C^1$ case. 

\begin{figure}[t]
\centering
\includegraphics[width=14cm]{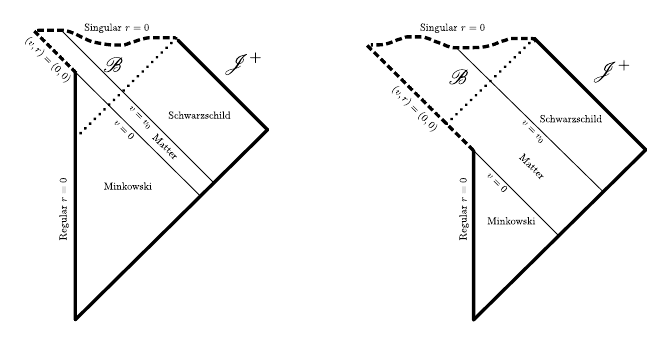}
\caption{ \footnotesize Penrose diagrams for two Vaidya spacetimes of interest, with smooth mass functions which grow at different rates and level off. The Minkowski, Schwarzschild, and matter-containing regions are labelled in each case, separated by the lines of $v=0$ and $v=v_0$. The dashed lines mark the singularities, while the dotted line marks the event horizon, above which is the black hole region $\mathscr{B}$. When the mass function increases more slowly (on the right), the $(v,r)=(0,0)$ singularity, which is stretched into a diagonal line in null-null coordinates, becomes globally naked. See Propositions \ref{proposition:locallynaked} and \ref{proposition:globallynaked}.}
\label{fig:vaidya1}
\end{figure}

For $v \leq 0$, the metric (\ref{equation:metric}) is precisely the Minkowski metric in ingoing null-radial coordinates. If $m(v)$ levels off to some fixed value $m_0$ at some $v_0>0$, i.e.\@ if $m(v) = m_0$ for $v \geq v_0$, then in this region (\ref{equation:metric}) is precisely the Schwarzschild metric of mass $m_0$ in ingoing Eddington-Finkelstein coordinates, so this is the sense in which $(M,g)$ is a simple model of the formation of a Schwarzschild black hole due to matter falling into the origin along the null geodesics of constant $0 < v < v_0$. While this is a useful image to keep in mind, the analysis will not require $m(v)$ to level off (so there need not be an exactly Schwarzschild region). See Figure \ref{fig:vaidya1} for a schematic Penrose diagram of two contrasting cases. There and in what follows, we suppress the $S^2$ coordinates.

A case which has received much attention is the self-similar case (\cite{dwivedi1989nature, dwivedi1991nature}), that of taking $m(v) = \frac{m_0}{v_0} v$ for $v \in [0,v_0]$ and $m(v) = m_0$ for $v \geq v_0$, and it is well understood that in this scenario a globally naked singularity develops at $(v,r) = (0,0)$ provided that $\frac{m_0}{v_0} \leq \frac{1}{16}$. In fact, it is true that a globally naked singularity develops under the significantly more general constraint that $\sup_{v>0} \frac{m(v)}{v} < \frac{1}{16}$, regardless of the form of $m$.  To show this, we simply investigate the ODE for outgoing radial null curves (which will necessarily be pre-geodesics): parameterizing a curve $\gamma$ by $v$ in $(v,r)$-coordinates, i.e.\@ taking $\gamma(v) = (v,r(v))$, we find that $\gamma'(v) = \partial_v + r'(v) \partial_r$ is null if and only if

\begin{equation}\label{equation:ode}
r'(v) = \frac{1}{2} - \frac{m(v)}{r(v)}.
\end{equation}
We would like to identify solutions to this ODE which terminate to the past at $r=0$ as $v \to 0$ (from above). Observe that the Picard-Lindel\"{o}f theorem ensures existence and uniqueness on $(v,r) \in \mathbb{R} \times \mathbb{R}_+$, and maximal solutions either exist for all $v \in \mathbb{R}$ or have $r \to 0$ at finite $v$.

Consider the curve $(v,\alpha(v))$ with $\alpha(v):= 2 m(v)$ (this is the apparent horizon), and define $\Delta_2 := r(v)-\alpha(v)$ for a solution $r(v)$ to the ODE (\ref{equation:ode}). Then we have
$$\Delta_2'(v) = r'(v)-2m'(v) \leq r'(v)  = \frac{1}{2}-\frac{m(v)}{r(v)} = \frac{\Delta_2(v)}{2r(v)},$$
so $\Delta_2$ will remain negative if it is ever negative. This is the familiar statement that the apparent horizon ``traps" the outgoing null curves. Hence if $r(v_0) \geq 2m(v_0) = \alpha(v_0)$ at some $v_0 > 0$, we must have had $r(v) \geq \alpha(v)$ for $v < v_0$.

We investigate when we can obtain similar comparisons via the curves associated to $\beta_k(v) := k m(v)$ with $k>0$. Define $\Delta_k(v) := r(v) - \beta_k(v)$ and consider 
\begin{align}\label{equation:deltaprime}
\Delta_k'(v) & = r'(v)-km'(v) = \frac{1}{2}-\frac{m(v)}{r(v)}-km'(v) \nonumber
\\ & = \left[\frac{1}{2} - \frac{1}{k} - km'(v)\right] + \frac{1}{k}-\frac{m(v)}{r(v)} \nonumber
\\ & = k\left[\frac{1}{2k} - \frac{1}{k^2} - m'(v)\right] + \frac{\Delta_k(v)}{kr(v)},
\end{align}
so $\Delta_k(v)$ will have a conserved sign provided that the term in brackets has fixed sign. This will be the key tool used in the following results. Defining $f: \mathbb{R}_+ \to \mathbb{R}$ by 
$$f(k) := \frac{1}{2k} - \frac{1}{k^2},$$
we see that the significance of the value $\frac{1}{16}$ is that it is the maximum of $f$, achieved at $k=4$, so controlling the comparison of $m'(v)$ to $\frac{1}{16}$ allows us to control the sign of $\Delta_k$.

We first treat the locally naked case.

\begin{proposition}\label{proposition:locallynaked}
(Locally Naked) Consider the incoming Vaidya spacetime as above characterized by a nondecreasing mass function $m(v)$ which is continuous and piecewise $C^1$. 
\begin{enumerate}[(i)]
\item If there exists a $v_0>0$ such that $m'(v) \leq \frac{1}{16}$ on $(0,v_0]$, then there exists a one-parameter family (modulo $S^2$) of outgoing radial null geodesics which terminate to the past at $(v,r) \to (0,0)$. 
\item If there exists a $v_0 > 0$ and $\varepsilon>0$ such that $m'(v) \geq \frac{1}{16} + \varepsilon$ on $(0,v_0]$, then no causal curve terminates to the past at $(v,r) \to (0,0)$.
\end{enumerate}
\end{proposition}

Note the hypothesis for {\it (i)} is ensured if $m_+'(0) < \frac{1}{16}$, while the hypothesis for {\it (ii)} is equivalent to $m_+'(0) > \frac{1}{16}$. It is already known by other techniques that the singularity is locally naked if and only if $m_+'(0) \leq \frac{1}{16}$ (\cite{bengtsson2012spherical, joshi1992strong}), but we include this statement and proof for use in and comparison to the globally naked result, and because the proof method makes trivial an observation regarding the curvature strength of the singularity, leading us to the next result.  In the globally $C^2$ case, of course, $m_+'(0) = m_-'(0)=0$, yielding the interesting observation that {\it smoothness favors a locally naked singularity}.

\proof {\it (i)}: Suppose $m'(v) \leq \frac{1}{16}$ on $(0,v_0]$. By the intermediate value theorem, we may find a $k_+ \in [4,\infty)$ such that $f(k_+) = \sup_{0 < v \leq v_0} m'(v)$. Consider a solution $r(v)$ to the ODE (\ref{equation:ode}) with initial value $0 < r(v_0) \leq k_+ m(v_0)$. By (\ref{equation:deltaprime}) and the hypothesis, 
$$\Delta_{k_+}'(v) \geq \frac{\Delta_{k_+}(v)}{k_+ r(v)}$$
on $(0,v_0]$, so $\Delta_{k_+}$ will remain positive on this domain if it is ever positive. Since $\Delta_{k_+}(v_0)$ is nonpositive by the initial condition, it must therefore be nonpositive for $0< v < v_0$. 

Since $m(v_0) > 0$, we may find a $0 < k_0 \leq 2$ such that $r(v_0) > k_0 m(v_0) = \beta_{k_0}(v_0)$, and (\ref{equation:deltaprime}) yields
$$\Delta_{k_0}'(v) \leq \frac{\Delta_{k_0}(v)}{k_0 r(v)},$$
so $\Delta_{k_0}$ remains negative if it is ever negative. Hence $\Delta_{k_0}$ is nonnegative for $0< v < v_0$. Putting these comparisons together, we've found
\begin{equation}\label{equation:compare}
k_0 m(v) \leq r(v) \leq k_+ m(v)
\end{equation}
for $0< v<v_0$, which ensures $ \lim_{v \to 0^+} r(v) = 0$, as desired.

{\it (ii)}: As in the proof of {\it (i)}, given any solution $r(v)$ to (\ref{equation:ode}), regardless of initial condition, we may find a $0 < k_0 \leq 2$ such that $\Delta_{k_0}\geq 0 $ for $v>0$ sufficiently small, which ensures that $(v,r(v))$ cannot terminate to the past at $(v,0)$ for any $v > 0$. Since solutions also cannot terminate to the future at $(v,0)$ for any $v \leq 0$, this further ensures that any maximal solution to (\ref{equation:ode}) has a value at some point in $(0,v_0]$, so initializing in this range of $v$ treats all outgoing radial null geodesics.

Now, suppose $m'(v) \geq \frac{1}{16}+ \varepsilon$ on $(0,v_0]$. We first note that since (\ref{equation:ode}) immediately yields that $r'(v) < \frac{1}{2}$, any solution which terminates to the past at $(0,0)$ must always satisfy the crude bound $r(v) < \frac{v}{2}$. We show that every solution somewhere violates this bound under our hypothesis. In particular, the hypothesis indicates that $m(v) > \frac{v}{16}$ on $(0,v_0]$, so it suffices to show that every solution satisfies $r(v) \geq 8 m(v)$ for some $v \in (0,v_0]$. 

Consider a solution $r(v)$ satisfying $r(v_*) < 8m(v_*)$ at some $v_* \in (0,v_0]$. We define the quantity $k(v)$ by $r(v) = k(v)m(v)$, so the initial condition stipulates $0 < k(v_*) < 8$, and we wish to show $k(v) \geq 8$ at some $v \in (0,v_0]$. Differentiating this definition yields
$$\frac{1}{2}- \frac{1}{k(v)} = r'(v) = k(v)m'(v)+k'(v)m(v),$$
which implies $k(v)$ satisfies
\begin{align} \label{equation:kode}
k'(v) & = \frac{ f(k(v)) - m'(v) }{m(v)} k(v)
\\ & \leq -\frac{\varepsilon}{m(v)}k(v) \nonumber
\end{align}
on $(0,v_0]$. The piecewise $C^1$ condition ensures we may also bound $m(v) \leq Cv$ on $(0,v_0]$, for some $C > 0$. Combining this with the above inequality gives
$$k'(v) \leq -\frac{\varepsilon}{C} \frac{k(v)}{v}$$
 on $(0,v_0]$, and an application of Gronwall's inequality now yields
$$k(v) \geq k(v_*) \left( \frac{v_*}{v} \right)^{\frac{\varepsilon}{C}}$$
for $v \in (0, v_*]$. This implies $k(v) \to \infty$ as $v \to 0$, and in particular that there is a $v \in (0, v_*]$ at which $k(v) \geq 8$. This demonstrates that there are no outward radial null geodesics which terminate to the past at $(0,0)$. 

If there were a radial causal curve $\gamma(s) := (v(s),r(s))$ which terminated to the past at $(0,0)$, then $\partial(J^+(\gamma))$ would be generated by outward radial null geodesics doing the same, so this cannot occur.  Since projecting a non-radial such causal curve to the $(v,r)$ coordinates yields a radial causal curve of the above form, these also cannot occur.

\qed

We observe that it is a trivial consequence of the proof of Proposition \ref{proposition:locallynaked}{\it (i)} that the Kretschmann scalar $K= \|\text{Riem}\|^2 = \frac{48m(v)^2}{r^6}$ diverges along all identified null geodesics as they limit to $(0,0)$, by virtue of the comparison $r(v) \leq k_+ m(v)$ in (\ref{equation:compare}) for $0< v \leq v_0$. We can show something stronger:

\begin{proposition}\label{proposition:kretschmann}
Consider the incoming Vaidya spacetime as above characterized by a nondecreasing mass function $m(v)$ which is continuous and piecewise $C^1$. There can exist at most one (modulo $S^2$) outgoing radial null geodesic which terminates to the past at $(v,r) \to (0,0)$ and along which the Kretschmann scalar $K = \frac{48m(v)^2}{r^6}$ does not diverge.
\end{proposition}

While such divergences were already known for the case $m_+'(0) > 0$ \cite{joshi1992strong}, it seems they were not known in general for $m_+'(0) = 0$. This significantly generalizes an explicit computation done by Joshi and Dwivedi \cite{joshi1992strong} which showed such divergences for a particular form of the mass function with $m_+'(0) = 0$ ($m(v) \sim v^n$, $n>1$). The present result indicates that the naked singularity in question {\it always} has divergent curvature associated with it, regardless of the form of $m(v)$. Importantly, this rules out the possibility of objecting to the physical significance of previously known results by maintaining that the singularity's strength was an artifact of a lack of regularity in the chosen mass function. Curvatures diverge no matter how smooth one takes $m(v)$.

\proof
We consider a solution $r(v)$ to (\ref{equation:ode}) satisfying $\lim_{v \to 0^+} r(v) = 0$, and as in the previous proof we consider the quantity $k(v)$ defined by $r(v) = k(v) m(v)$. We note that, along the curve of interest,
$$K(v) = \frac{48m(v)^2}{r(v)^6} = \frac{48}{r(v)^4 k(v)^2},$$
so $\lim_{v \to 0^+} K(v) = \infty$ provided that $k(v)$ is bounded as $v \to 0$. It is not difficult to see from the bound $r(v) < \frac{v}{2}$ that $k(v)$ is bounded near $v = 0$ if $m_+'(0) > 0$ (meaning $K$ always diverges in this scenario), so assume $m_+'(0) = 0$. It can be seen in this case that if $k(v)$ is unbounded near $v=0$, then in fact $\lim_{v \to 0^+} k(v) = \infty$: if not, there is an $C > 0$ such that $4 < k(v_n) < C$ along some positive sequence $(v_n)_{n=1}^\infty$ with $v_n \to 0$, so eventually $\sup_{0<v<v_n} m'(v) < f(C) < f(k(v_n)) $, which implies by (\ref{equation:kode}) that $k(v) \leq k(v_n)$ for $v \in (0,v_n]$, showing that $k(v)$ is bounded near $v=0$.

We now assume there exist two such solutions $r_1(v)$, $r_2(v)$ to (\ref{equation:ode}), satisfying $\lim_{v \to 0^+} r_i(v) = 0$ and $\lim_{v \to 0^+} k_i(v) = \infty$ for $i=1,2$, and show they must be equal. In particular, for a fixed $\varepsilon > 0$ there is a $v_0 > 0$ such that $r_i(v) > (4+\varepsilon)m(v)$ for $v \in (0,v_0]$-- this is all that will be needed from the $k_i(v) \to \infty$ hypothesis to deduce the equality. We restrict attention to the interval $(0,v_0]$. Setting $\lambda := \frac{1}{2}-\frac{1}{4+\varepsilon}$, we have by (\ref{equation:ode}) that $\lambda v < r_i(v) < \frac{v}{2}$, so we further have that $\frac{r_2}{r_1} > 2\lambda$. Observe

\begin{align}\label{equation:ricompare}
(r_1-r_2)' & = \frac{m}{r_2}-\frac{m}{r_1} = \frac{m}{r_2} \left(1-\frac{r_2}{r_1} \right) \nonumber
\\ & < \frac{1-\frac{r_2}{r_1} }{4+\varepsilon}.
\end{align}

Setting $C_0 := \frac{1-2\lambda}{4+\varepsilon} = \frac{2}{(4+\varepsilon)^2}$, then, (\ref{equation:ricompare}) together with the aforementioned bound on $\frac{r_2}{r_1}$ indicates 
$$r_1(v)-r_2(v) < C_0v.$$
Dividing through by $r_1$ and rearranging, however, yields the new bound $\frac{r_2}{r_1} > 1-\frac{C_0}{\lambda}$. Combining this new bound with (\ref{equation:ricompare}) now yields
$$r_1(v)-r_2(v) < C_1 v, $$
where $C_1 := \frac{C_0}{\lambda(4+\varepsilon)} = \frac{2}{2+\varepsilon}C_0.$ Iterating this procedure, we deduce for each $n \in \mathbb{Z}_+$ that 
$$r_1(v)-r_2(v) < C_n v,$$
where $C_n := \left(\frac{2}{2+\varepsilon} \right)^n C_0$. Since $\lim_{n \to \infty} C_n = 0$ (and by symmetry between $r_1$ and $r_2$), this implies $r_1(v) = r_2(v)$, as desired.

\qed

We learn from this result and its proof that, while in the case $0 < m_+'(0) \leq \frac{1}{16}$ every outgoing radial null geodesic terminating to the past at $(v,r) \to (0,0)$ has bounded $k(v)$ (and hence divergent $K(v)$) near $v=0$, when $m_+'(0) = 0$ there is exactly one (modulo $S^2$) outgoing radial null geodesic terminating to the past at $(v,r) \to (0,0)$ along which $k(v)$ is unbounded near $v=0$, and this curve necessarily generates the Cauchy horizon associated to a complete spacelike hypersurface in $M$. This curve's $K(v)$ will behave in boundedness like $\frac{m(v)}{v^3}$ near $v=0$, which may be bounded or unbounded depending upon the precise mass function (it is bounded for $m$ sufficiently smooth). All others in the continuum of outgoing radial null geodesics emanating from $(0,0)$ will have bounded $k(v)$ (in fact, they will have $\lim_{v \to 0^+} k(v) = 2$), and hence divergent $K(v)$, near $v = 0$.

We now treat the globally naked subcase. 

\begin{proposition}\label{proposition:globallynaked}
(Globally Naked) Consider the incoming Vaidya spacetime as above characterized by a nondecreasing mass function $m(v)$ which is continuous and piecewise $C^1$.

\begin{enumerate}[(i)]
\item If $\sup_{v>0} \frac{m(v)}{v} < \frac{1}{16}$, then there exists a one-parameter family (modulo $S^2$) of outgoing radial null geodesics which both terminate to the past at $(v,r) \to (0,0)$ and reach $r \to \infty$ to the future.

\item If there exists a $v_0 > 0$ at which $\frac{m(v_0)}{v_0} \geq \frac{1}{4}$, then no causal curve both terminates to the past at $(v,r) \to (0,0)$ and reaches $r \to \infty$ to the future.
\end{enumerate}
\end{proposition}

We remark that the hypothesis of {\it (i)} essentially stipulates that the {\it average} growth rate of the mass function (between $0$ and $v$) is always less than $\frac{1}{16}$, while that of {\it (ii)} stipulates that the average growth is at some point at least $\frac{1}{4}$. The heuristic physical takeaway from part {\it (i)} of Propositions \ref{proposition:locallynaked} and \ref{proposition:globallynaked}, then, is that if the in-falling mass accumulates at the origin {\it slowly} enough, a signal from the initial singularity has time to escape before enough mass gathers to trap it. Part {\it (ii)} of these propositions indicates that these naked singularities {\it require} slow accumulation.

\proof {\it (i)}: Choose $C>0$ with $\sup_{v>0} \frac{m(v)}{v} < C < \frac{1}{16}$. Notice that, in particular, $m(v) < Cv$ for $v>0$. We first show the conclusion holds for $\widetilde m(v) = Cv$ (a result already well-known by explicit computation, but which we show here for completeness), then compare. Quantities below with a tilde are defined according to the mass function $\widetilde m(v)$ instead of $m(v)$.

Fix $v_0>0$. By the Intermediate Value Theorem, we may find $k_\pm$ with $k_- \in (2,4)$ and $k_+ \in (4,\infty)$ such that $f(k_\pm) = C$. Following the proof of Proposition \ref{proposition:locallynaked}{\it (i)}, we consider a solution $\tilde r(v)$ to (\ref{equation:ode}) satisfying $k_- \widetilde m(v_0) < \tilde r(v_0) \leq k_+ \widetilde m(v_0)$. (\ref{equation:deltaprime}) yields
$$\widetilde \Delta_{k_\pm}'(v) = \frac{\widetilde \Delta_{k_\pm}(v)}{k_\pm \tilde r(v)},$$
so $\widetilde \Delta_{k_-}(v)$ remains positive for all $v > v_0$ since it was positive initially, while $\widetilde \Delta_{k_+}(v)$ is nonpositive for all $0 < v < v_0$ since it was nonpositive initially. That is, 
\begin{align*}
\tilde r(v) & > k_- \widetilde m(v) = k_- Cv, \quad v \in [v_0,\infty) \\
\tilde r(v) & \leq k_+ \widetilde m(v) = k_+ Cv, \quad v \in (0,v_0]
\end{align*}
These inequalities imply $\tilde r(v)$ will terminate to the past at $(0,0)$ and reach $r \to \infty$ to the future.

Now, for each such $\tilde r(v)$ consider a solution $r(v)$ to (\ref{equation:ode}) (now with $m(v)$) satisfying $r(v_0) = \tilde r(v_0)$. Defining $\Delta(v) := r(v) - \tilde r (v)$, we have
\begin{align*}
\Delta'(v) & = r'(v) - \tilde r'(v)
\\ & = \frac{\widetilde m(v)}{\tilde r(v)} - \frac{m(v)}{r(v)}
\\ & > Cv \left[ \frac{1}{\tilde r(v)} - \frac{1}{r(v)} \right]
\\ & = \frac{Cv}{r(v) \tilde r(v)} \Delta(v).
\end{align*}
Hence $\Delta$ remains positive if it is ever positive, and if $\Delta$ is ever $0$, it must immediately become and remain positive. These constraints ensure that, since $\Delta(v_0) = 0$, we have $\Delta(v) > 0$ for $v > v_0$ and $\Delta(v) < 0$ for $v < v_0$, showing that $\lim_{v \to 0^+} r(v) = 0$ and $\lim_{v \to \infty} r(v) = \infty$ from the same limits for $\tilde r(v)$.

{\it (ii)}: Suppose $m(v_0) \geq \frac{v_0}{4}$ for some $v_0 > 0$. Consider a solution $r(v)$ to (\ref{equation:ode}), and observe that if it is to reach $r \to \infty$, we certainly must have $r(v_0) \geq 2m(v_0)$ (if not, as usual we have $\Delta_2 < 0$ for $v>v_0$, which implies by (\ref{equation:ode}) that $r'(v)$ is negative for $v>v_0$). Hence a solution which escapes to infinity satisfies
$$r(v_0) \geq 2m(v_0)  \geq \frac{v_0}{2},$$
which implies $r(v)$ cannot terminate to the past at $(0,0)$, as discussed in the proof of Proposition \ref{proposition:locallynaked}{\it (ii)}. Also as discussed there, this rules out any causal curve with the same properties.

\qed

The argument of ${\it (i)}$ for $\widetilde m(v) = Cv$ carries through nearly exactly for any $m(v)$ with $\sup_{v>0} m'(v) = C < \frac{1}{16}$, but the less restrictive hypothesis of $\sup_{v>0} \frac{m(v)}{v} < \frac{1}{16}$ is accommodated by comparing to the linear case. One may observe that we only needed the strict inequality in $\sup_{v} \frac{m(v)}{v} < \frac{1}{16}$ for $v > v_0$, so one could relax this to an inclusive inequality on $v < v_0$. If one stipulates that the mass function should level off, for example, one may relax the hypothesis to $\sup_{v} \frac{m(v)}{v} \leq \frac{1}{16}$. 

One may also observe that the hypotheses of ${\it (i)}$ and ${\it (ii)}$ above are, apart from the discrepancy between the values $\frac{1}{16}$ and $\frac{1}{4}$, very nearly logical negations of each other. This might lead us to hope that we may be able to close the gap between these values and obtain a complete characterization of globally naked singularities in these spacetimes. Marginal improvements as in the previous paragraph notwithstanding, this is unfortunately not possible in a direct manner: neither value can be improved, and one can either have or not have globally naked singularities in the gap. Regarding ${\it (i)}$, for any $\varepsilon > 0$ there exist mass functions with $\sup_{v > 0}  \frac{m(v)}{v} < \frac{1}{16}+\varepsilon$ which have no causal curves ${\it either}$ terminating at $(0,0)$ to the past ${\it or}$ reaching $r \to \infty$ to the future-- this is seen in the linear case $m(v) = Cv$ with $\frac{1}{16} < C < \frac{1}{16}+\varepsilon$. Regarding ${\it (ii)}$, for any $\varepsilon > 0$ there exist mass functions possessing a $v_0 > 0$ at which $\frac{m(v_0)}{v_0} \geq \frac{1}{4}-\varepsilon$ while still having outgoing radial null geodesics which both terminate to the past at $(0,0)$ and reach $r \to \infty$ to the future-- though some technical details must be worked through to show this, the underlying idea is to consider an $m(v)$ which closely approximates the step function $v_0(\frac{1}{4}-\frac{\varepsilon}{2}) H(v-v_0)$, where $H$ is the Heaviside step function, while satisfying our regularity and structural constraints. 

These considerations apparently indicate that Proposition \ref{proposition:globallynaked} is optimal inasmuch as the range of $\frac{m(v)}{v}$ alone can readily characterize globally naked singularities. Further results in this vein must therefore be structurally different, perhaps hypothesizing a constraint on the duration that $\frac{m(v)}{v}$ exceeds $\frac{1}{16}$ (notice that step functions minimize this duration subject to the monotonicity condition). In any case, Proposition \ref{proposition:globallynaked} is sufficient for the physical observations presently of interest.

\section{Conclusions}

Proposition \ref{proposition:globallynaked}{\it (i)} implies that the singularity at $(v,r) \to (0,0)$ is visible from infinity for an extended period provided, in particular, that the open condition $\sup_{v>0} m'(v) < \frac{1}{16}$ is met, while Proposition \ref{proposition:kretschmann} implies that this singularity is always physically strong in some sense. A reasonable topology one might put on the collection of incoming Vaidya spacetimes is that induced by putting them in bijection with the set of admissible $m(v)$ endowed with the $C^1$-type norm 
$$\|m\| = \sup_{v>0} \left[ |m_-'(v)| + |m_+'(v)| \right].$$
Notice that this topology has the minimal fineness in $m$, and therefore in the initial data induced on some complete spacelike hypersurface, required to capture ``closeness" in $G$ (given (\ref{equation:curvature})), and hence the implicit matter distribution. In such a reasonable topology, the subset of Vaidya spacetimes exhibiting globally naked singularities has nonempty interior. We note that, since $m(v)$ is not required to level off, this observation is true both in and outside of the asymptotically flat context: the hypothesis of Proposition \ref{proposition:globallynaked}{\it (i)} can be met even while $\lim_{v \to \infty} m(v) = \infty$. 

These observations are significant because they add to the small list of examples wherein naked singularities are apparently generic (at least, within spherical symmetry), this conclusion being made tractable due to the Vaidya spacetimes' relative simplicity-- compared, say, to Tolman-Bondi models. Though the genericness is within the class of Vaidya spacetimes, varying within this class does not seem meaningfully more restrictive than varying initial data in a manner consistent with a particular ``fundamental" matter field, in regards to degrees of freedom. One hopes for the sake of potential implications on weak cosmic censorship, of course, that this genericness is destroyed by perturbing outside of spherical symmetry. While the physical heuristic that slow accumulation allows an escaping signal certainly seems independent of spherical symmetry, it may well be that the singularity's forming early enough for it to be the source of such a signal depends critically on spherical symmetry's generically allowing the perfect focusing of matter shells at the origin. 

While the Vaidya spacetimes are clearly in violation of the physical spirit of the Weak Cosmic Censorship (WCC) Conjecture, they are apparently outside the technical scope of current rigorous formulations of the conjecture as an initial value problem (see, e.g.\@ \cite{christodoulou1999global, waldgr}), as they do not a priori stipulate a specific matter field coupling to the Einstein equation to yield a PDE governing how the spacetime can be evolved from initial data. Since these spacetimes satisfy the DEC, however, and since the complete suite of fundamental physical matter fields is not known, one might argue they have just as much claim to exhibiting physical behavior as do toy matter models. 

Of course, though the denomination is somewhat nebulous, we should not dismiss out of hand that there is a well-reasoned school of thought that one should restrict entirely to ``fundamental" matter models when formulating cosmic censorship so as to filter out singularities that may arise purely out of approximations and idealizations of matter (e.g.\@ pressureless dust). This perspective can at least be traced to Eardley and Smarr \cite{eardley1979time} (who credit an unpublished report by Hawking), and further discussion can be found in Chapter 12 of Wald \cite{waldgr}. Much of the discussion on this topic predates Christodoulou's work (\cite{christodoulou1994examples, christodoulou1999instability}) on the scalar field case demonstrating that a ``generic" qualifier is unavoidable, however, which opened up the question of whether both this {\it and} a restriction to fundamental matter is necessary. That is, it is worthwhile to recognize the logical distinction between WCC formulations stating 
\begin{enumerate}[(i)]
\item Spacetimes subject to the DEC do not admit globally naked singularities.
\item Spacetimes containing only fundamental matter do not admit globally naked singularities.
\item Spacetimes subject to the DEC do not {\it generically} admit globally naked singularities.
\item Spacetimes containing only fundamental matter do not {\it generically} admit globally naked singularities.
\end{enumerate}
While examples demonstrated very early on that (i) could not be true, Christodoulou's work in the 1990s demonstrated that (ii) is almost certainly not true either, though (iii) or (iv) may well be. Knowing which, if either, of these options ends up being viable would provide significant insight into the structure and content of general relativity. To put the present work into the context of this conversation, our results provide some of the strongest evidence that we are aware of (but still not definitive due to the restriction to spherical symmetry) that (iii) may not be true, while making no direct implications about (iv) (though hints surrounding the structure of naked singularities gleaned herein may be helpful in understanding (iv)).

In any event, it is clear from the manner in which our (and many others') analysis has proceeded that the physical evaluation of cosmic censorship in examples is often most naturally done in regards to the totality of a spacetime and its structure, rather than through initial data and evolutions thereof. In a forthcoming publication, we put forward a more comprehensive technical formulation of WCC which is better aligned with this perspective, and hence more naturally accommodates the physically objectionable behavior demonstrated here-- that is, a technical formulation better tuned to handling the broadest interpretation of version (iii) above.

\section{Acknowledgements}

The author would like to extend gratitude to his advisor, Dr.\@ Redacted1, for supporting this work, as well as Dr.\@ Redacted2 for providing thoughtful discussion on closely related problems to those treated here. In particular, the method of Proposition \ref{proposition:kretschmann}'s uniqueness proof is largely due to Dr. Redacted2, adapted from application to a related ODE. We are also grateful to the referees for their constructive input.

\bibliographystyle{plain}
{\small \bibliography{references.bib}}

\end{document}